\documentclass[preprint,prb,amsmath,superscriptaddress,aps,subeqn,showpacs]{revtex4}
\usepackage{graphicx}

\begin{document}
\title{Surface Polar Phonon Dominated Electron Transport in Graphene}

\author{X. Li}
\affiliation{Department of Electrical and Computer Engineering, North Carolina State University, Raleigh, NC 27695-7911}

\author{E. A. Barry}
\affiliation{Weapons and Materials Research Directorate, US Army Research Laboratory, Aberdeen Proving Grounds, Maryland, 21005-5069}

\author{J. M. Zavada}
\affiliation{Department of Electrical and Computer Engineering,
North Carolina State University, Raleigh, NC 27695-7911}

\author{M. Buongiorno Nardelli}
\affiliation{Department of Physics, North Carolina State University,
Raleigh, NC 27695-8202}
\affiliation{CSMD, Oak Ridge National
Laboratory, Oak Ridge, TN 37831}

\author{K. W. Kim}
\affiliation{Department of Electrical and Computer Engineering,
North Carolina State University, Raleigh, NC 27695-7911}


\begin{abstract}
The effects of surface polar phonons on electronic transport properties of monolayer graphene are studied by using a Monte Carlo simulation. Specifically, the low-field electron mobility and saturation velocity are examined for different substrates (SiC, SiO$_{2}$, and HfO$_{2}$) in comparison to the intrinsic case.  While the results show that the low-field mobility can be substantially reduced by the introduction of surface polar phonon scattering, corresponding degradation of the saturation velocity is not observed for all three substrates at room temperature.  It is also found that surface polar phonons can influence graphene electrical resistivity even at low temperature, leading potentially to inaccurate estimation of the acoustic phonon deformation potential constant.

\end{abstract}

\maketitle

Since the experimental realization of graphene in 2004,~\cite{Novoselov2004} its extraordinary electric properties have led to near exponentially growing attnetion in possible applications. In particular, the linear band structure in the vicinity of the Dirac points, and resultant massless fermions, have excited interest in both its applicability for electronic devices and as a testbed for quantum field theory phenomena.\cite{CastroNeto2009} The extremely high intrinsic mobilities that have been observed in suspended graphene at room temperature,\cite{Bolotin2008} in conjunction with its true two dimensional structure of graphene, naturally lead to its employment in higher speed graphene transistors devoid of short channel effects.\cite{Schwierz2010} For such high frequency applications, the saturation current and low-field mobility are material properties that need to be accurately determined.

When a graphene layer is in close proximity to a polar substrate, inelastic carrier scattering with surface polar phonons (SPPs) can result in significant reduction in the low-field mobility of graphene.~\cite{Chen2008,Fratini2008,Konar2009,Perebeinos2010} Due to the inelastic nature of SPP, they also provide pathway to current saturation, in conjunction with, or as a substitute for, intrinsic optical phonons.  While a number of studies generally suggested the negative influence of reduced saturation velocities (due to the relative small energies of SPPs),\cite{Meric2008,Chauhan2009,Barreiro2009} conflicting reports exist that predict a very different picture (including enhanced velocities) based on the analysis of Boltzmann transport equation.~\cite{Perebeinos2010,Ashley2010}

In this work, we investigate the influence of carrier scattering due to SPPs on the electronic transport properties of monolayer graphene, via a full-band ensemble Monte Carlo method. Particularly, the low-field electron mobility and saturation velocity are calculated in the presence of three different substrates (SiC, SiO$_{2}$, and HfO$_{2}$) and compared with those of intrinsic graphene at room temperature.  Furthermore, we examine the impact of SPPs on the low temperature electrical resistivity, with attention to its implication on the experimental determination of the acoustic phonon deformation potential constant.

The Monte Carlo model adopted in the calculation utilizes the complete electron and phonon spectra in the first Brillouin zone.  While a tight-binding band is used for the electronic energy structure, all six branches of the graphene phonon spectra are considered with the phonon dispersion and electron-phonon scattering rates obtained from \textit{first principles} calculations.~\cite{Borysenko2010} The effect of degeneracy is accounted for by the rejection technique, after final state selection.~\cite{Li2010} The distribution function is obtained {\em self-consistently} from the ensemble simulation.  The SPP scattering rate is introduced by following~\cite{Fratini2008,Konar2009}
\begin{equation}
\label{SPPRate}
\frac{1}{\tau_{S}(k_{i})}=\frac{2\pi}{\hbar}\sum_{q}\frac{e^{2}\mathcal{F}^{2}}
{2\varepsilon(q)^{2}}\frac{e^{-2qd}}{q}(1+\cos\theta)(n_q+
\frac{1}{2}\pm\frac{1}{2})\delta(E_f-E_i\pm\hbar\omega_{S}) \,,
\end{equation}
where $q=|\bf{k_f}-\bf{k_i}|$ is the SPP momentum, $E_f$ ($E_i$) the final (initial) electron energies, $d$ the distance between monolayer graphene and the substrate (0.4 nm), $\omega_{S}$ the SPP energy, and  $\mathcal{F}^2=\frac{\hbar\omega_{S}}{2A\varepsilon_0} (\frac{1}{\kappa_{S}^{\infty}+1}-\frac{1}{\kappa_{S}^{0}+1})$ the square of Fr\"{o}hlich coupling constant.  The $(1+\cos\theta)$ factor originates from the overlap integral of the pseudospin part of the electron wavefunction and $\varepsilon(q)$ is the dielectric function in graphene. In addition, $\mathcal{F}$ contains the dependence on high (low) frequency dielectric constant of the polar substrate $\kappa^{\infty}_{S}$ ($\kappa^{0}_{S}$) along with the normalized area $A$.  The scattering by ionized impurities in the substrate is not included in the effort to clearly identify the role of SPPs.  The values for relevant substrate parameters can be found in Refs.~\onlinecite{Perebeinos2010,Fratini2008,Konar2009}.

Figure~1 shows the SPP scattering rates calculated at 300~K for three substrates, SiC, SiO$_2$, and HfO$_2$.  For comparison, the strength of electron-optical phonon interaction inherent in graphene (via the deformation potential) is also plotted from a first-principles analysis.~\cite{Borysenko2010}  The results clearly illustrate the dominance of SPPs over intrinsic optical phonons in graphene for the entire electron energy under consideration.  Furthermore, this enhancement in scattering is more pronounced for the substrates with small $\omega_S$ that is apparent from the onset energy of emission process (e.g., HfO$_2$  - 19.4~meV; SiO$_2$ - 60.0~meV; SiC - 116~meV).  Due to the Coulombic nature, the SPP scattering is a function of electron density $n$ and subsequent screening in graphene. Throughout the calculation, we assume $ n= 1 \times 10^{12}$~cm$^{-2}$ along with the static screening function $\varepsilon(q)$ in the random-phase approximation.~\cite{Hwang2007}  The intrinsic scattering via deformation potential interaction has no dependence on $n$.

Figure~2 shows the electron drift velocity vs.\ the electric field at T$=$300 K, for intrinsic graphene and graphene on different substrates.  As expected, the addition of electron-SPP interactions lead to general decrease of electron drift velocities in the low-field region.  Consequently, the low-field mobilities are reduced for all three substrate with the largest decrease in HfO$_2$.  However, the difference with the intrinsic case becomes progressively smaller with the increasing field (thus, the increasing average electron energy) and, in the case of SiO$_2$ and SiC, the velocity appears to saturate at a higher value.  No degradation is observed even for HfO$_2$.  Clearly, this behavior does not follow the saturation velocity model suggested by Ref.~\onlinecite{Meric2008} based on the onset of optical phonon emission, intrinsic phonon or otherwise (i.e., SPPs).  Rather, it is in general agreement with more recent studies that solved the Boltzmann transport equation numerically assuming a displaced Fermi-Dirac distribution function.~\cite{Perebeinos2010,Ashley2010}

The apparent inconsistency of increased total scattering rate and absence of velocity degradation in the high-field region may be explained by considering the linear energy dispersion and the characteristics of the Fr\"{o}hlich interactions.  As it is well known, a change in electron energy (say, via an inelastic scattering) does not automatically relax the drift velocity in monolayer graphene near the $K$ and $K^\prime$ points.  What matters is the direction of the final momentum.  Due to the Coulombic nature,  the Fr\"{o}hlich interactions, including those by SPPs, prefer small angle events.
When a SPP scattering occurs, the electron will likely emit (absorb) a SPP, with the phonon momentum pointing along the radial direction toward (away from) the Dirac point. Consequently, the velocity of the scattered electron may not change significantly.  In contrast, optical phonon scattering of intrinsic graphene is via deformation potential interactions and, thus, randomize/relax effectively the direction of the electron momentum/drift velocity.  Ignoring the difference between the Fr\"{o}hlich and deformation potential interactions~\cite{Chauhan2009} can lead to inaccurate depiction of transport properties.

When the electron-SPP interactions are taken into account along with other scattering mechanisms, they provide an additional channel for energy and momentum relaxation.  Accordingly, the average electron energy becomes substantially lower at a given electric field.  As the applied field increases, it means that the shift in the distribution function (along the direction of the electric field) would be smaller in energy along with a shorter tail compared to that in intrinsic graphene.  This is clearly visible in Fig.~3 plotted for the $k_y=0$ cross-section at 20 kV/cm. The observed shorter tails in the negative $k_x$ space (i.e., with negative velocities) as well as in the high energy region with nonlinear dispersion~\cite{Perebeinos2010} appear to more than compensate the additional momentum relaxation of SPP scattering on the SiO$_2$ and SiC substrates, leading to larger saturation velocities shown in Fig.~2.  In the case of HfO$_2$, the distribution is much less heated with a significant population in the negative half space.  Due to the very strong inelastic scattering by surface phonons, the drift velocity may not have reached the saturation point at 20 kV/cm and could grow further.  Cooling of electron distributions via surface phonon interactions may also reduce the self-heating as the generated heat (i.e., SPPs) is on the substrate and, thus, can be more readily dissipated.  Another interesting point to note from Fig.~3 is that the electron distribution function in graphene does not resemble that of a highly degenerate case, certainly not when subject to an appreciable electric field.  A simple approximation of a displaced box-like function~\cite{Meric2008,Barreiro2009} cannot describe transport properties accurately.


The strong influence of the substrate on low-field mobility observed at 300~K raises a possibility that the SPP scattering may be an efficient mechanism even at low temperatures.  This is a distinct possibility due to the small phonon energies $\omega_S$.  Figure~4 plots the electrical resistivity $\rho$ in graphene vs.\ temperature  for all four cases under consideration.  Indeed, the resistivity is substantially affected by the substrate conditions at $T \gtrsim 50$~K and the effect of SPP scattering can not be immediately separated from those of other mechanisms (most importantly, the acoustic phonon scattering).  The acoustic deformation potential is often determined experimentally utilizing the assumption that the low temperature resistivity is linearly dependent on temperature with a slope that is proportional to the square of acoustic deformation potential.~\cite{Chen2008,Hwang2008}  However, the calculation results indicate that this may not be possible; the extracted deformation potential ($D_{ac}$) is not an intrinsic quantity but an "effective" parameter that includes substrate effects and screening (i.e., the charge density in graphene) among others.

Concerning the specific results, the intrinsic case gives $D_{ac} \approx 6.8$~eV with a well-defined linear region.  This (i.e., the strong linearity) is due to the negligible contribution of optical phonon scattering at low temperatures with relatively large energies ($\hbar \omega_{op} \sim $ 160~meV).  The small deviation of $D_{ac}$ from the deformation potential constant obtained directly from the acoustic phonon scattering rate ($ \approx 4.5$~eV)~\cite{Borysenko2010} may be attributed to a degenerate electron density considered in the present calculation ($ n= 1 \times 10^{12}$~cm$^{-2}$).  It is also interesting to note that this value (6.8~eV) is actually rather close to the experimentally extracted on a non-polar substrate (7.8~eV).~\cite{Hong2009}  For the SiC substrate,  we estimate $D_{ac}\approx 7.1$~eV similar to the intrinsic result.  As $T\gtrsim 150$~K, however, there is a substantial increase in SPP scattering and the difference between the SiC and intrinsic cases becomes more discernible.  When SiO$_2$ is used as the substrate, the slope of $\rho$ is further increased and becomes nonlinear earlier due to the small SPP energy.  The deduced value in the linear region ($T\lesssim 125$~K) gives the appearance of $D_{ac}\approx 13.2$~eV, which is not unlike 16$-$18~eV estimated experimentally on SiO$_2$.~\cite{Chen2008}  Finally, HfO$_2$ has the largest effect among the three substrates.  The resistivity is greatly increased with no apparent linear region in the temperature range under consideration.  Accordingly, it is not plausible to determine $D_{ac}$.  The relevance of SPP scattering even in the $T \lesssim 150$~K regime can explain, at least in part, the very disparate results for the magnitude of acoustic phonon deformation potential in  graphene.~\cite{Chen2008,Perebeinos2010,Borysenko2010,Hong2009}

In summary, we investigate the effects of substrate-induced SPP scattering on the electronic transport properties in monolayer graphene by using a full-band ensemble Monte Carlo simulation.  It is found that the electron velocity-field characteristics are highly dependent on the choice of substrate at room temperature.  Specifically, the Fr\"{o}hlich nature of the interaction appears to be crucial for correctly describing the saturation of drift velocity.  The simulation also shows that the SPP scattering remains an efficient mechanism even at low temperatures ($T \gtrsim 50$~K), substantially affecting the electrical resistivity.  This makes it difficult to experimentally determine the acoustic deformation potential for intrinsic graphene in close proximity to a substrate.

This work was supported, in part, by the Defense Advanced Research Projects Agency/HRL Laboratories (the CERA program), the US Army Research Offce, and the SRC Focus Center on Functional Engineered Nano Architectonics (FENA). MBN wishes to acknowledge partial support from the Offce of Basic Energy Sciences, US DOE at Oak Ridge National Lab under contract DE-AC05-00OR22725 with UT-Battelle, LLC. JMZ acknowledges support from NSF under the IR/D program.

\clearpage
\noindent Figure Captions

\vspace{0.5cm} \noindent Figure~1. (Color online) Surface polar phonon scattering rate of graphene electron on SiC, SiO$_2$,
and HfO$_2$, for the electron density $ n$ of $ 1 \times 10^{12}$~cm$^{-2}$ at 300~K.  Also plotted is the intrinsic graphene optical phonon scattering rate at 300~K ("intrinsic") obtained from Ref.~\onlinecite{Borysenko2010}.

\vspace{0.5cm} \noindent Figure~2. (Color online) Electron drift velocity in graphene on SiC, SiO$_2$, and HfO$_2$.  The intrinsic case without substrate is also shown.  The electron density is $1 \times 10^{12}$~cm$^{-2}$ at 300~K.

\vspace{0.5cm} \noindent Figure~3. (Color online) Electron distribution function with different substrate conditions:  intrinsic (dash-dotted), SiC (solid), SiO$_2$ (dashed), and HfO$_2$ (dotted), for the $k_y=0$ cross-section at 20 kV/cm. The box-like function corresponds to the Fermi-Dirac distribution displaced by the SiO$_2$ SPP energy (60 meV) in a simple, metal-like approximation (i.e., $E_F \gg k_B T$) with $ n = 1 \times 10^{12}$~cm$^{-2}$.  For comparison, the equilibrium Fermi-Dirac distribution at 300~K is also plotted.

\vspace{0.5cm} \noindent Figure~4. (Color online) Electrical resistivity vs.\ temperature with different substrate conditions: intrinsic (circle), SiC (square), SiO$_2$ (triangle), and HfO$_2$ (diamond), with $ n = 1 \times 10^{12}$~cm$^{-2}$.  The slope of the straight lines are used to extract the acoustic deformation potential.

\clearpage
\begin{center}
\begin{figure}
\includegraphics[bb=243 514 416 672]{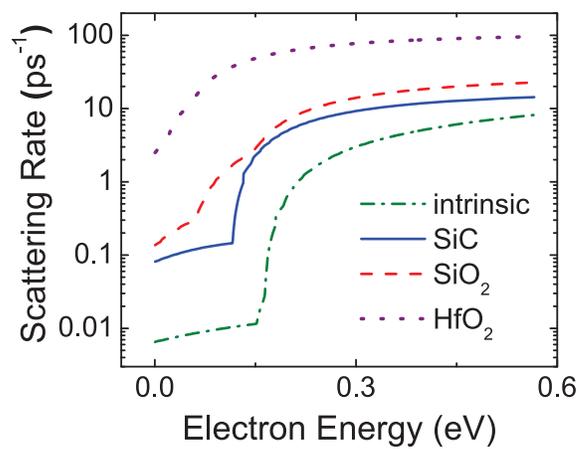}
\caption{{\large Li {\em et al.}} }
\end{figure}
\end{center}

\clearpage
\begin{center}
\begin{figure}
\includegraphics[bb=243 514 416 672]{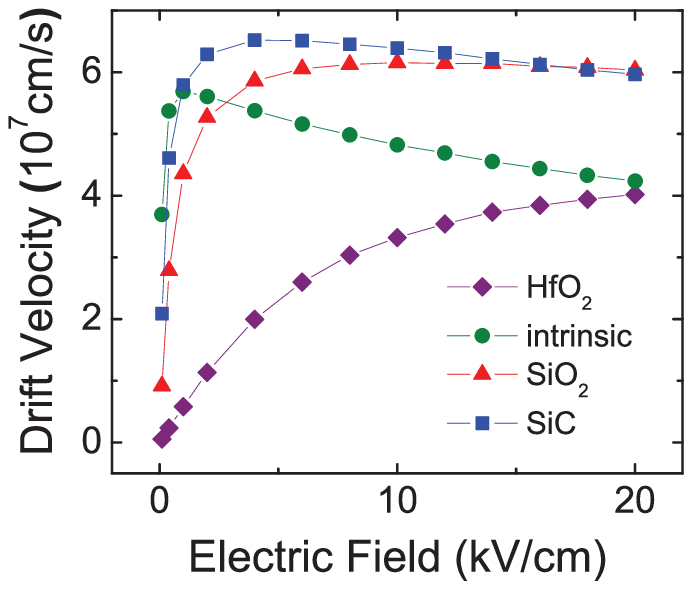}
\caption{{\large Li {\em et al.}} }
\end{figure}
\end{center}

\clearpage
\begin{center}
\begin{figure}
\includegraphics[bb=243 514 416 672]{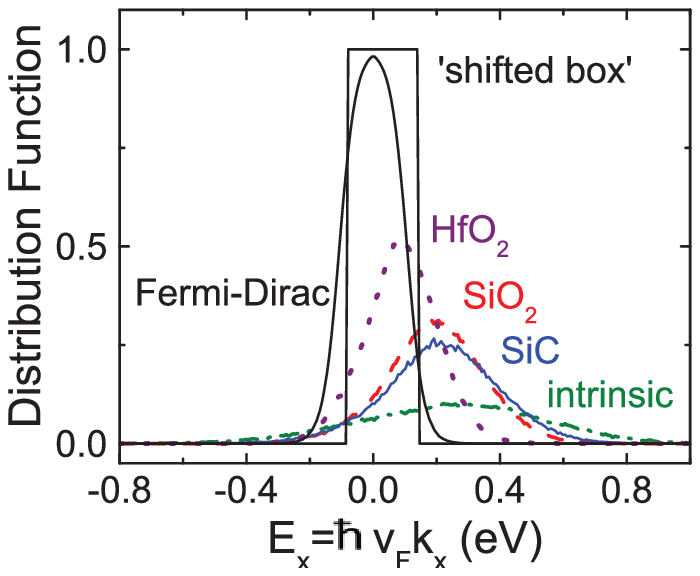}
\caption{{\large Li {\em et al.}} }
\end{figure}
\end{center}

\clearpage
\begin{center}
\begin{figure}
\includegraphics[bb=243 514 416 672]{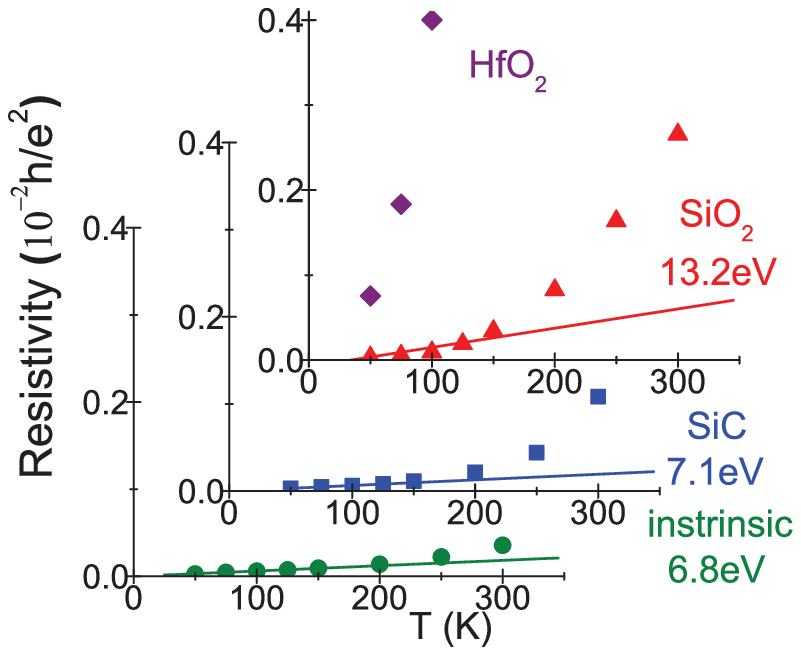}
\caption{{\large Li {\em et al.}} }
\end{figure}
\end{center}

\end{document}